\documentclass[aps,prl,showpacs,superscriptaddress,longbibliography,twocolumn,floatfix]{revtex4-1}
\usepackage{graphicx}
\usepackage{amsmath}
\usepackage{amssymb}
\usepackage{mathtools}
\usepackage{makeidx}
\usepackage{amsfonts}
\usepackage{bm}
\usepackage{dcolumn}
\usepackage{color}
\usepackage{xcolor}
\usepackage{xfrac}
\usepackage{wasysym}
\usepackage[colorlinks,
linkcolor=blue,
anchorcolor=blue,
citecolor=blue,
urlcolor=blue
]{hyperref}
\usepackage[normalem]{ulem}

\graphicspath{{./figures/}}

\begin{document}

\newcommand{\nwc}{\newcommand}
\nwc{\vs}{\vspace}
\nwc{\hs}{\hspace}
\nwc{\la}{\langle}
\nwc{\ra}{\rangle}
\nwc{\nn}{\nonumber}
\nwc{\Ra}{\Rightarrow}
\nwc{\wt}{\widetilde}
\nwc{\lw}{\linewidth}
\nwc{\ft}{\frametitle}
\nwc{\ben}{\begin{enumerate}}
\nwc{\een}{\end{enumerate}}
\nwc{\bit}{\begin{itemize}}
\nwc{\eit}{\end{itemize}}
\nwc{\dg}{\dagger}
\nwc{\mA}{\mathcal A}
\nwc{\mD}{\mathcal D}
\nwc{\mB}{\mathcal B}

\nwc{\Tr}[1]{\underset{#1}{\mbox{Tr}}~}
\nwc{\D}[2]{\frac{d #1}{d #2}}
\nwc{\pd}[2]{\frac{\partial #1}{\partial #2}}
\nwc{\ppd}[2]{\frac{\partial^2 #1}{\partial #2^2}}
\nwc{\fd}[2]{\frac{\delta #1}{\delta #2}}
\nwc{\pr}[2]{K(i_{#1},\alpha_{#1}|i_{#2},\alpha_{#2})}
\nwc{\av}[1]{\left< #1\right>}
\nwc{\alert}[1]{\textcolor{red}{#1}}
\nwc{\alertans}[1]{\textcolor{blue}{#1}}

\nwc{\zprl}[3]{Phys. Rev. Lett. ~{\bf #1},~#2~(#3)}
\nwc{\zpre}[3]{Phys. Rev. E ~{\bf #1},~#2~(#3)}
\nwc{\zpra}[3]{Phys. Rev. A ~{\bf #1},~#2~(#3)}
\nwc{\zjsm}[3]{J. Stat. Mech. ~{\bf #1},~#2~(#3)}
\nwc{\zepjb}[3]{Eur. Phys. J. B ~{\bf #1},~#2~(#3)}
\nwc{\zrmp}[3]{Rev. Mod. Phys. ~{\bf #1},~#2~(#3)}
\nwc{\zepl}[3]{Europhys. Lett. ~{\bf #1},~#2~(#3)}
\nwc{\zjsp}[3]{J. Stat. Phys. ~{\bf #1},~#2~(#3)}
\nwc{\zptps}[3]{Prog. Theor. Phys. Suppl. ~{\bf #1},~#2~(#3)}
\nwc{\zpt}[3]{Physics Today ~{\bf #1},~#2~(#3)}
\nwc{\zap}[3]{Adv. Phys. ~{\bf #1},~#2~(#3)}
\nwc{\zjpcm}[3]{J. Phys. Condens. Matter ~{\bf #1},~#2~(#3)}
\nwc{\zjpa}[3]{J. Phys. A ~{\bf #1},~#2~(#3)}
\nwc{\zpjp}[3]{Pramana J. Phys. ~{\bf #1},~#2~(#3)}

\title{Thermodynamic Trade-off Relation for First Passage Time in Resetting Process}
\author{P. S. Pal}
\affiliation{School of Physics, Korea Institute for Advanced Study, Seoul 02455, Korea}
\author{Arnab Pal}
\affiliation{The Institute of Mathematical Sciences, CIT Campus, Taramani, Chennai 600113, India \& Homi Bhabha National Institute, Training School Complex, Anushakti Nagar, Mumbai 400094, India}
\author{Hyunggyu Park}
\affiliation{Quantum Universe Center, Korea Institute for Advanced Study, Seoul 02455, Korea}
\author{Jae Sung Lee}
\affiliation{School of Physics, Korea Institute for Advanced Study, Seoul 02455, Korea}
\begin{abstract}
Resetting is a strategy for boosting the speed of a target-searching process. Since its introduction over a decade ago, most studies have been carried out under the assumption that resetting takes place instantaneously. However, due to its irreversible nature, resetting processes incur a thermodynamic cost, which becomes infinite in case of instantaneous resetting.  Here, we take into consideration both the cost and first passage time (FPT) required for a resetting process, in which the reset or return to the initial location is implemented using a trapping potential over a finite but random time period. An iterative generating function and counting functional method à la Feynman and Kac are employed to calculate the FPT and average work for this process. From these results, we obtain an explicit form of the time-cost trade-off relation, which provides the lower bound of the mean FPT for a given work input when the trapping potential is linear. This trade-off relation clearly shows that instantaneous resetting is achievable only when an infinite amount of work is provided. More surprisingly, the trade-off relation derived from the linear potential seems to be valid for a wide range of trapping potentials. In addition, we have also shown that the fixed-time or sharp resetting can further enhance the trade-off relation compared to that of the stochastic resetting.
\end{abstract}
\maketitle

\section{Introduction}  Reset refers to a process that completely erases information pertaining to the current state of a system and returns the system to a predetermined setting. Due to the irreversible nature of the erasure process, a certain thermodynamic cost is required for physical implementation of the reset. According to Landauer's principle, the minimum cost for information erasure is $k_{B}T \ln 2$ of dissipated heat to reset one bit of information~\cite{Landauer, Parrondo}, where $k_{B}$ is the Boltzmann constant and $T$ is the environmental temperature. This minimum bound is attainable only in a quasi-static process, which takes an infinitely long time. To reduce the reset time, additional cost should be incurred~\cite{Diana,Boyd,Schmiedl,Proesmans1,Proesmans2,Berut,Jun}. This indicates the existence of a time-cost trade-off for the reset process, implying that less energy consumption requires more time and vice versa; this trade-off fundamentally constrains the performance of the reset process~\cite{Zhen, Lee}. It will be demonstrated that the trade-off relation prevents instantaneous reset (zero reset time) unless an infinite amount of energy is provided, which is not feasible in the real world. Consequently, study on the minimal time of a process accompanied with resetting should
take into account the thermodynamic trade-off relations which has been a prominent topic in the field of stochastic thermodynamics
for the past decade~\cite{Shiraishi_power, Lee_origin, Barato, Dechant, Horowitz, Hasegawa, Lee_universal, Shirarish_speed, Falasco,Ito, Lee_underdamped}.

\begin{figure}
	\centering
	\includegraphics[width=0.45\textwidth]{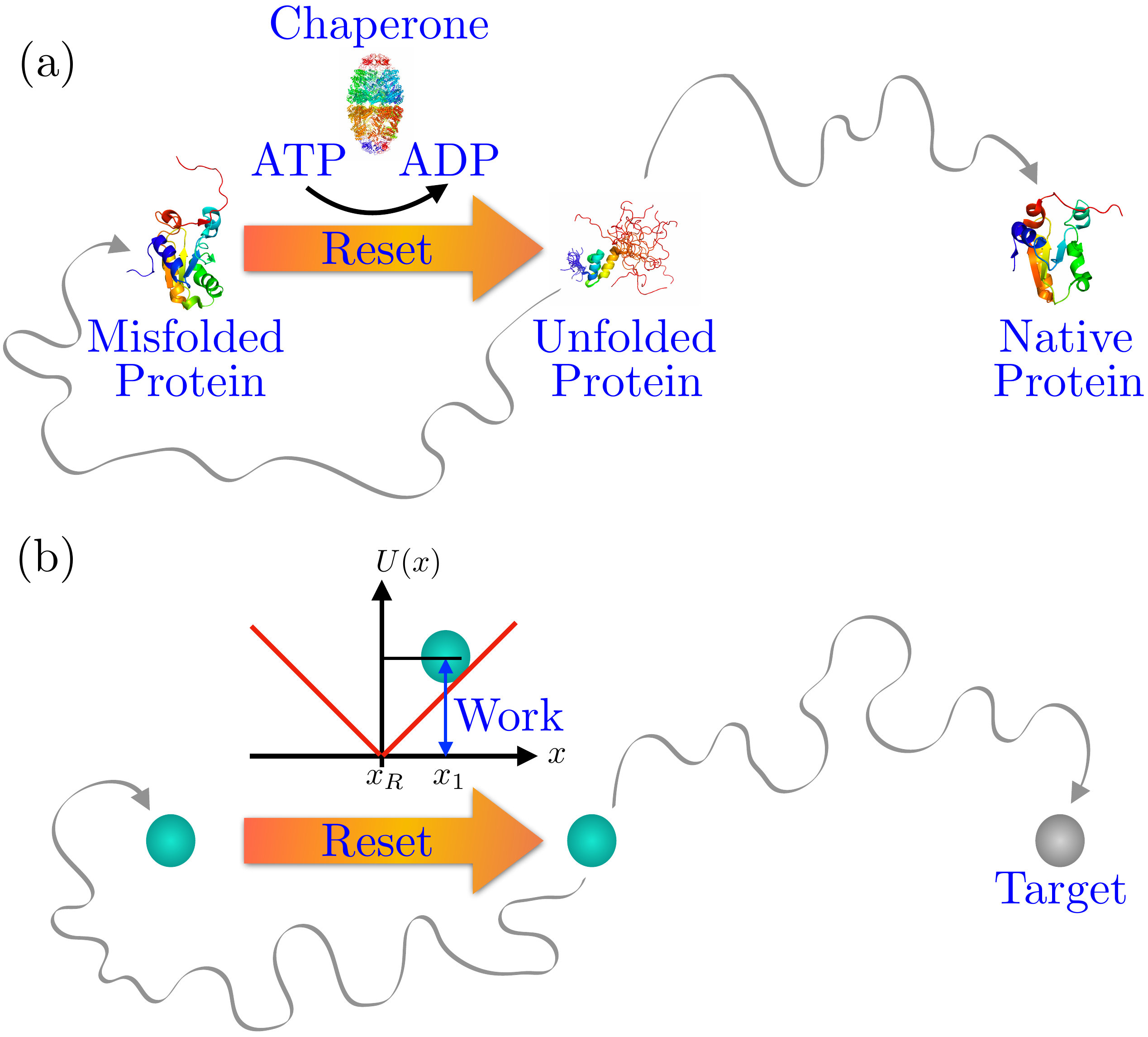}
	\caption{Schematics of finite-time stochastic resetting processes.  (a) Chaperone-assisted protein folding dynamics, from an initial unfolded to final native state,  in a rugged free energy landscape. When the protein is trapped in a local minima (misfolded state), a chaperone assists the protein to be reset to the initial unfolded state. (b) Stochastic resetting run by a Brownian particle. At a random time with a fixed rate, the position of the particle is reset by applying an external potential centered around the reset position.}
	\label{fig:schematic}
\end{figure}

Interestingly, it has recently been verified that resetting is an important mechanism to boost the speed of target searching using a random walker~\cite{evans2011, gupta2014, majumdar2015, evans2019,eule2015, pal-ness,mendez2016, basu2019, reuveni2016, pal2017, pal2019prl, kusmirez2014, chechkin2018, belan2018}. This target-searching strategy, referred to as {\em stochastic resetting}, is crucial for speeding up several biological processes such as kinetic proofreading~\cite{BarZiv, Murugan}, the chaperone-assisted protein-folding process~\cite{Bhaskaran, Hyeon, Chakrabarti}, molecular transport \cite{FJ}, and chemical reaction \cite{unbinding}. For example, protein-folding dynamics can be viewed as a random walk construct, which starts from an initial unfolded state and then searches for the target (native state) in a rugged free-energy landscape, as shown in Fig.~\ref{fig:schematic}(a). During the process, proteins are sometimes trapped in a local minima (misfolded state), which significantly prolongs the search time. The chaperone assists in restoring misfolded proteins back to the initial unfolded state. Then, the search begins again. This reset significantly reduces the target-searching time. However, the ATP hydrolysis is necessary for chaperone-assisted resetting of the misfolded protein. This example clearly shows that the time for searching, accompanied by stochastic resetting, should be limited by the thermodynamic cost.

Previous studies on stochastic resetting focused mainly on the search time when the reset is carried out instantaneously, without consideration for the cost. A more proper question regarding stochastic resetting in reality should be
``What is the minimum search time for \emph{a given limited energetic cost}?'', which is the main subject of this study. To answer this, we consider a \emph{finite-time} reset process implemented by a trapping potential~\cite{pal2019njp,pal2019pre, maso2019, bodrova2020pre1, bodrova2020pre2,pal2020prr,st-ret,st-ret-1} and evaluate the total work for reset and the global first passage time (FPT), which includes the time for reset.
Based on this result, we derive a time-cost trade-off relation for stochastic resetting. Although this trade-off relation is derived only for a linear potential case, we numerically demonstrate its validity for a wide range of trapping potentials. In addition, we show that the trade-off relation can be further enhanced when the resets occur after fixed time intervals.


\section{Setup}  We consider a stochastic resetting process for a one-dimensional Brownian particle in an overdamped environment, as shown in Fig.~\ref{fig:schematic}(b). The particle undergoes free-diffusion motion (with diffusion constant $D$) in diffusion phase starting from an initial position $x_0$. At a random time drawn from an exponential distribution $f_R(t)=re^{-rt}$ where $r$ is the rate, a potential $U(x)$ is turned ON in order to bring the particle to a predetermined reset position $x_{\rm R}$. The potential is maintained ON until the particle reaches $x_{\rm R}$ with a different diffusion constant $D_{\rm R}$ (the reset or return phase). As soon as the particle reaches $x_{\rm R}$, the potential is turned OFF and the particle resumes its free diffusion motion. This diffusion and resetting process is repeated until the particle reaches or finds the target position at $x_{\rm T}$ during its diffusive phase~\cite{mis}.
We set $x_{\rm T}$ to the origin without loss of generality. This process corresponds to A process  in Fig.~\ref{Fig:ABCprocess} and can be described by the following Langevin equations:
\begin{align}
	\dot x =
	\left\{\begin{array}{ll}
		\sqrt{2D}\eta(t), & \text{ diffusion phase} \\
		-\partial_x U(x)+\sqrt{2D_{\rm R}}\eta(t) , & \text{ reset phase }
	\end{array} \right. \label{eq:model_eq}
\end{align}
where $\eta(t)$ is a Gaussian white noise with zero mean and unit variance. The global FPT $t_{\rm G}$ is given by $t_{\rm G} = t_{\rm D} + t_{\rm R}$ (see A process in Fig.~\ref{Fig:ABCprocess}), where $t_{\rm D}$ and $t_{\rm R}$ are the time spent in the diffusion and reset phases until the particle reaches the target, respectively. We note that $t_{\rm D}$ corresponds to the FPT when the reset is instantaneous (see C process in Fig.~\ref{Fig:ABCprocess}). Such stochastic resetting process with stochastic
returns using external trap was introduced in \cite{st-ret,st-ret-1} where various non-equilibrium properties such as the steady state and relaxation phenomena were studied.
Evidently, our model~\eqref{eq:model_eq} is a generalization of those in Ref.~\cite{bodrova2020pre1, pal2020prr}, where only deterministic dynamics of the reset phase, i.e. $D_{\rm R}=0$, was considered.  

\section{First passage time} 
\begin{figure}
	\centering
	\includegraphics[width=0.38\textwidth]{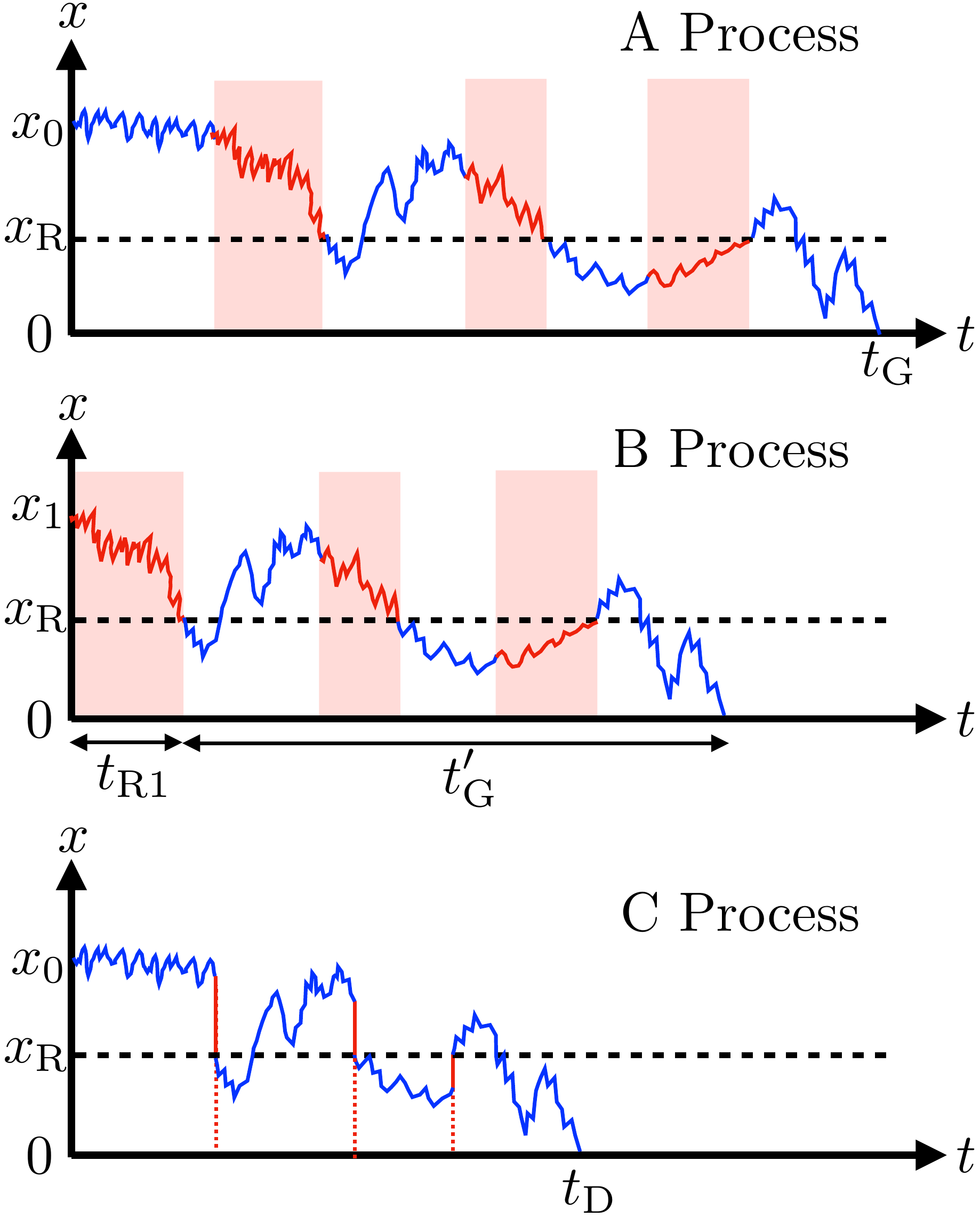}
	\caption{Stochastic resetting trajectories of A, B, and C processes. A process starts from $x_0$ in the diffusion phase, and then, sequential reset (red-shaded region) and diffusion (unshaded region) phases continue until the particle touches the target.
The FPT for the A process is given by $t_{\rm G}=t_{\rm D}+t_{\rm R}$.  B process starts from $x_1$ in the reset phase with duration $t_{\rm R1}$, and then, sequential diffusion and reset phases continue until the particle touches the target with duration $t_{\rm G}^\prime$. C process is obtained by eliminating all the reset phases of the A process, thus the FPT of the C process is simply given by $t_{\rm D}$.}
	\label{Fig:ABCprocess}
\end{figure}
Inspired by the Brownian functional method~\cite{curr-sc,Singh}, we introduce the iterative generating function method to evaluate the $n$th moment of $t_{\rm G}$.  First, the moment generating function of $t_{\rm G}$ for the A process (starting in the diffusion phase) can be written as
\begin{align}
	Q_{\rm A} (p|x_{0}) =\int_{0}^{\infty}dt_{\rm G} ~e^{-p t_{\rm G}} P_{\rm A} (t_{\rm G}|x_{0}) \equiv \langle e^{-p t_{\rm G}}  \rangle,
\end{align}
where $P_{\rm A}(t_{\rm G}|x)$ is the probability density function of $t_{\rm G}$ for the A process with the initial position $x$.

Now we consider the B process as illustrated in Fig.~\ref{Fig:ABCprocess}; it starts from $x_1$ in the reset phase and terminates when the particle touches the target (origin) in the diffusion phase. The B process can be divided into two parts;
the first reset phase and the remaining part with durations $t_{\rm R1}$ and $t_{\rm G}^\prime$, respectively. Then, the moment generating function of the global FPT for the B process, $t_{\rm R1}+t_{\rm G}^\prime$, can be written as
\begin{align}
	Q_{\rm B} (p|x_{1}) =\int_{0}^{\infty} dt_{\rm R1} \int_{0}^{\infty} dt_{\rm G}^\prime ~e^{-p(t_{\rm R1} + t_{\rm G}^\prime)} P_{\rm B} (t_{\rm R1},t_{\rm G}^\prime|x_{1}) ,
\end{align}
where $P_{\rm B} (t_{\rm R1},t_{\rm G}^\prime|x_1)$ is the joint probability density function of $t_{\rm R1}$ and $t_{\rm G}^\prime$ for the B process with the initial position $x_1$. As $t_{\rm R1}$ and $t_{\rm G}^\prime$ are independent variables, $P_{\rm B} (t_{\rm R1},t_{\rm G}^\prime|x_{1}) = P_{\rm R} (t_{\rm R1}| x_1) P_{\rm A} (t_{\rm G}^\prime| x_{\rm R})$, where $P_{\rm R} (t| x)$ is the probability density function for a single reset phase taking time $t$ with the initial position $x$. Consequently,
$Q_{\rm B}(p|x_1)$ can be expressed in a product form as
\begin{align}
	Q_{\rm B} (p|x_{1}) = Q_{\rm R} (p|x_{1}) Q_{\rm A} (p|x_{\rm R}),  \label{eq:Q_B}
\end{align}
where $Q_{\rm R} (p|x_{1})$ is the momentum generating function of the FPT for a single reset phase defined as
\begin{align}
	Q_{\rm R} (p|x) =\int_{0}^{\infty}dt ~e^{-p t} P_{\rm R} (t|x). \label{eq:gen_reset}
\end{align}
See the Appendix A for detailed calculation of $Q_{\rm R} (p|x)$.

Now we construct a differential equation for $Q_{\rm A} (p|x_0)$ with respect to $x_0$.
We divide the A process with duration $t_G$ into the initial infinitesimal part with duration $\Delta t$ and the remaining part with duration $t_G-\Delta t$. During the first infinitesimal diffusion process, the dynamics may be switched into the reset phase with probability $r\Delta t$ (reset rate $r$), and then the remaining process becomes the B process starting from $x_0^\prime\equiv x(\Delta t)=x_0 + \sqrt{2D}\eta(0)\Delta t$.
Otherwise, the diffusion phase continues with probability $1-r\Delta t$ and then the remaining one becomes the A process starting from $x_0^\prime$.
Therefore,  one can write $Q_{\rm A}(p|x_0)$ in an iterative way as
\begin{align}
	Q_{\rm A} (p|x_{0}) &=  e^{-p \Delta t}  \left\langle e^{-p(t_{\rm G} - \Delta t )}   \right\rangle \nonumber \\
	&=  e^{-p \Delta t}  \left\langle Q_{\rm A}(p|x_0^\prime)(1-r\Delta t) + Q_{\rm B}(p|x_0^\prime) r\Delta t   \right\rangle .\label{eq:Q_A_iterative}
\end{align}
Equation \eqref{eq:Q_A_iterative} is iterative in $Q_{\rm A}(p|x)$, since $Q_{\rm B}(p,x_0^\prime)$ can be replaced by $Q_{\rm A}(p|x)$ from Eq.~\eqref{eq:Q_B}.
By expanding Eq.~\eqref{eq:Q_A_iterative} and keeping terms up to the order of $\Delta t$, one can obtain the following backward differential equation of $Q_{\rm A}(p|x_0)$:
\begin{align}
	\left[D\partial_{x_0}^2 -(p+r) \right] Q_{\rm A}(p|x_0)+rQ_{\rm R}(p|x_0) Q_{\rm A}(p|x_R) = 0.
	\label{bde_gt_d}
\end{align}
By solving Eq.~\eqref{bde_gt_d}, one can obtain $Q_{\rm A}(p|x_0)$. Then, the $m$th moment of $t_{\rm G}$ can be calculated as
\begin{align}
	\la t_{\rm G}^m\ra=(-1)^m\partial_p^mQ_{\rm A}(p|x_0)\bigg|_{p\rightarrow 0}.
	\label{A_moments}
\end{align}

\section{Work for reset} 
Work fluctuations in resetting processes have been of topical interest in recent times (see \cite{work-1,work-2}). However, all these works measure work due to the modulation of an external control parameter. In contrast, in our set-up, work is done when the external potential is switched on at the beginning of every reset phase and thus the Brownian particle gains energy from the external potential. Since the potential has no time dependence during the reset phase, there is no further work done on the particle. Note that the potential energy gained by the particle is completely dissipated as heat throughout the reset phase.  Suppose there is a single stochastic trajectory, as shown in the A process of Fig.~\ref{Fig:ABCprocess}, where each reset phase starts at time $t_i$ ($i=1,2,\cdots$). The total work $W$ for the trajectory is calculated as the sum of the potential values evaluated at $t_i$, with the potential set to zero at the reset point. This summation can be expressed more generally by the \textit{counting functional}:
\begin{align}
	V[x(t)]=\int_0^{t_{\rm G}} Z[x(t)] dt,  \label{eq:counting_functional}
\end{align}
where $Z[x(t)]= \sum_i\delta(t-t_i)w(x(t))$ with a weight function $w(x(t))$ evaluated at $t$. This functional yields the quantities related to the number of resets during the whole process: For instance, when $w(x)=1$, $V[x(t)]$ yields the total number of resets during the  process and, when $w(x)=U(x)$, $V[x(t)]$ corresponds to the total work $W$.

Evaluating the general moments of $V$ can be conveniently carried out by considering the trajectory of the C process as shown in Fig.~\ref{Fig:ABCprocess}, which is obtained by eliminating all reset phases from the original trajectory of the A process. It is important to note that the counting functional $V$ yields the same value for both trajectories.  In fact, the trajectory of the C process corresponds to that of \emph{instantaneous} resetting. Thus, Eq.~\eqref{eq:counting_functional} can be evaluated on the corresponding trajectory of the C process with $V[x(t)]=\int_0^{t_{\rm D}} Z[x(t)]dt$. The generating function for $V$ is then given as
\begin{align}
	Q_{\rm C} (p|x_{0}) =\int_{0}^{\infty}dV ~e^{-p V} P_{\rm C} (V|x_{0}) \equiv \langle e^{-p V} \rangle, \label{eq:moment_gen_func_C}
\end{align}
where $P_{\rm C} (V|x_{0})$ is the probability density function of $V$ for the C process. By dividing $V[x(t)]$ into the initial infinitesimal part with duration $\Delta t$ and the remaining part as $V[x(t)] = Z[x(0)] \Delta t + \int_{\Delta t}^{t_{\rm D}} Z[x(t)]dt$, one can rewrite Eq.~\eqref{eq:moment_gen_func_C} as
\begin{align}
	Q_{\rm C} (p|x_{0}) =   \left\langle  e^{-p Z[x(0)] \Delta t} e^{-p\int_{\Delta t}^{t_{\rm D}} Z[x(t)] dt }  \right\rangle .\label{eq:Q_C_expansion1}
\end{align}
During the initial infinitesimal process, resetting occurs with probability $r\Delta t$. Then the next position $x_0^\prime$ is reset to  $x_{\rm R}$ and $Z[x(0)]\Delta t = w(x_0)$. Otherwise, with probability $1-r\Delta t$, the diffusion phase continues, thus, $x_0^\prime = x_0 + \sqrt{2D}\eta(0)\Delta t$  and $Z[x(0)]\Delta t = 0$. Similar to Eq.~\eqref{eq:Q_A_iterative},  Eq.~\eqref{eq:Q_C_expansion1} can be written as
\begin{align}
	Q_{\rm C} (p|x_{0}) =&  (1-r\Delta t)  \left\langle Q_{\rm C}(p|x_0+\sqrt{2D}\eta(0)\Delta t) \right\rangle \nonumber \\
	&+  r\Delta t e^{-p w(x_0)} Q_{\rm C}(p|x_{\rm R}) .
	 \label{eq:Q_C_expansion2}
\end{align}
By keeping terms up to the order of $\Delta t$, we finally arrive at the following backward differential equation of $Q_{\rm C}(p|x_0)$:
\begin{align}
	\left[D\partial^2_{x_0}-r \right] Q_{\rm C}(p|x_0)+re^{-pw(x_0)}Q_{\rm C}(p|x_R)=0.
	\label{eq:bde_C}
\end{align}
The $m$th moment of $V$ is then calculated as
\begin{align}
	\la V^m\ra=(-1)^m\partial_p^m Q_{\rm C}(p|x_0)\bigg|_{p\rightarrow 0}.
	\label{C_moments}
\end{align}

\section{Linear potential case} 
For the sake of simplicity, $D$ is set to $\frac{1}{2}$ from now on.
We analytically solve the backward differential equations~\eqref{bde_gt} and \eqref{eq:bde_C} for a linear potential in $x$ as
\begin{align}
	U(x) = a|x-x_{\rm R}|^n \label{eq:pot}
\end{align}
with $a>0$ and $n=1$. This linear potential leads to $Q_{\rm R}(p|x) = e^{-\lambda (p) |x-x_{\rm R}|}$, with $\lambda(p)=(\sqrt{a^2+4pD_{\rm R}}-a)/2D_{\rm R}$. This $Q_{\rm R}(p|x)$ can be used to solve Eq.~\eqref{bde_gt_d} explicitly (see Appendix B). In particular, for $x_0 = x_{\rm R}$, the solution of Eq.~\eqref{bde_gt_d} is rather simpler as
\begin{align}
Q_{\rm A}(p|x_R)=\frac{e^{-\mu(p)x_{\rm R}}}{f_{\rm A}(p,x_{\rm R})}, \label{eq:sol_QA}
\end{align}
where $\mu(p) = \sqrt{(p+r)/D}$ and  $f_{\rm A} (p,x_{\rm R})=1+\nu(p)-\nu(p)e^{-[\lambda(p)+\mu(p)]x_{\rm R}}-\frac{2\nu(p)\lambda(p)}{\mu(p)}\sinh[\mu(p)x_{\rm R}]e^{-\mu(p)x_{\rm R}}$ with $\nu(p) = r/[D\lambda^2(p)- (p+r)]$. The mean FPT is then evaluated by utilizing Eq.~\eqref{A_moments}. The result is
\begin{align}
\la t_{\rm G}\ra&=-\partial_p Q_{\rm A}(p|x_{\rm R})\bigg|_{p\rightarrow 0} \nn\\
&=\la t_{\rm D}\ra+\frac{1}{a\alpha}[2\sinh(\alpha x_{\rm R})-\alpha x_{\rm R}],
\label{fpt}
\end{align}
where $\alpha=\sqrt{r/D}$ and $\la t_{\rm D}\ra=(e^{\alpha x_{\rm R}}-1)/r$ is the mean FPT with the instantaneous resetting strategy~\cite{evans2011}. The detailed derivations for Eqs.~\eqref{eq:sol_QA} and \eqref{fpt} are provided in Appendix B. The second term of the right-hand side in Eq.~\eqref{fpt} represents the (positive) extra  time due to our finite-time reset strategy. As $a\rightarrow\infty$ (infinitely steep potential), this extra time vanishes, thus representing the limit of instantaneous resetting. 

To evaluate the total work, we substitute $w(x(t))=a|x-x_{\rm R}|$ into Eq.~\eqref{eq:bde_C} and solve for 
$Q_{\rm C}(p|x_0)$ with $x_0=x_{\rm R}$. We find 
\begin{align}
	Q_{\rm C}(p|x_{\rm R})=\frac{e^{-\alpha x_{\rm R}}}{f_{\rm C}(p,x_{\rm R})}, \label{eq:Q_C_linear_pot}
\end{align}
where $f_{\rm C}(p,x_{\rm R})=1+\gamma (p)-\gamma (p)e^{-(pa+\alpha)x_{\rm R}}-\frac{2pa}{\alpha}\gamma (p)\sinh(\alpha x_{\rm R})e^{-\alpha x_{\rm R}}$ with $\gamma(p) = r/ (Da^2p^2 - r) $. The mean total work is then calculated as
\begin{align}
	\la W\ra&=-\partial_pQ_{\rm C}(p|x_{\rm R})\bigg|_{p\rightarrow 0} \nonumber \\
	&=\frac{a}{\alpha}[2\sinh(\alpha x_{\rm R})-\alpha x_{\rm R}].
	\label{work1}
\end{align}
Note that the work diverges in the $a\rightarrow \infty$ limit (instantaneous resetting), indicating an intrinsic trade-off between time and cost.
The detailed derivation for Eqs.~\eqref{eq:Q_C_linear_pot} and \eqref{work1} are provided in Appendix C. We now turn our attention to another interesting limit $r \rightarrow 0$ (thus, $\alpha \rightarrow 0$). Intuitively, one would expect the average work to be zero in this limit since no work is done without resetting. However, following Eq.~(\ref{work1}), one finds $\lim_{r \rightarrow 0} \la W\ra =  ax_{\rm R}$, which is a finite non-zero quantity that depends on the potential strength. This implies that the limit of zero resetting rate is not equivalent to the bare process where no reset takes place at all. Thus, there exists a discontinuity in the average work at $r=0$. This is attributed to the fact that average number of resets vanishes as $\alpha x_{\rm R}$, but the work per one reset diverges as $a/\alpha$ since the particle diffuses far away from the origin for a long duration before reset occurs. For this reason, mean work remains to be finite in the limit $r \to 0$.

\begin{figure}[h]
	\centering
	\includegraphics[width=0.45\textwidth]{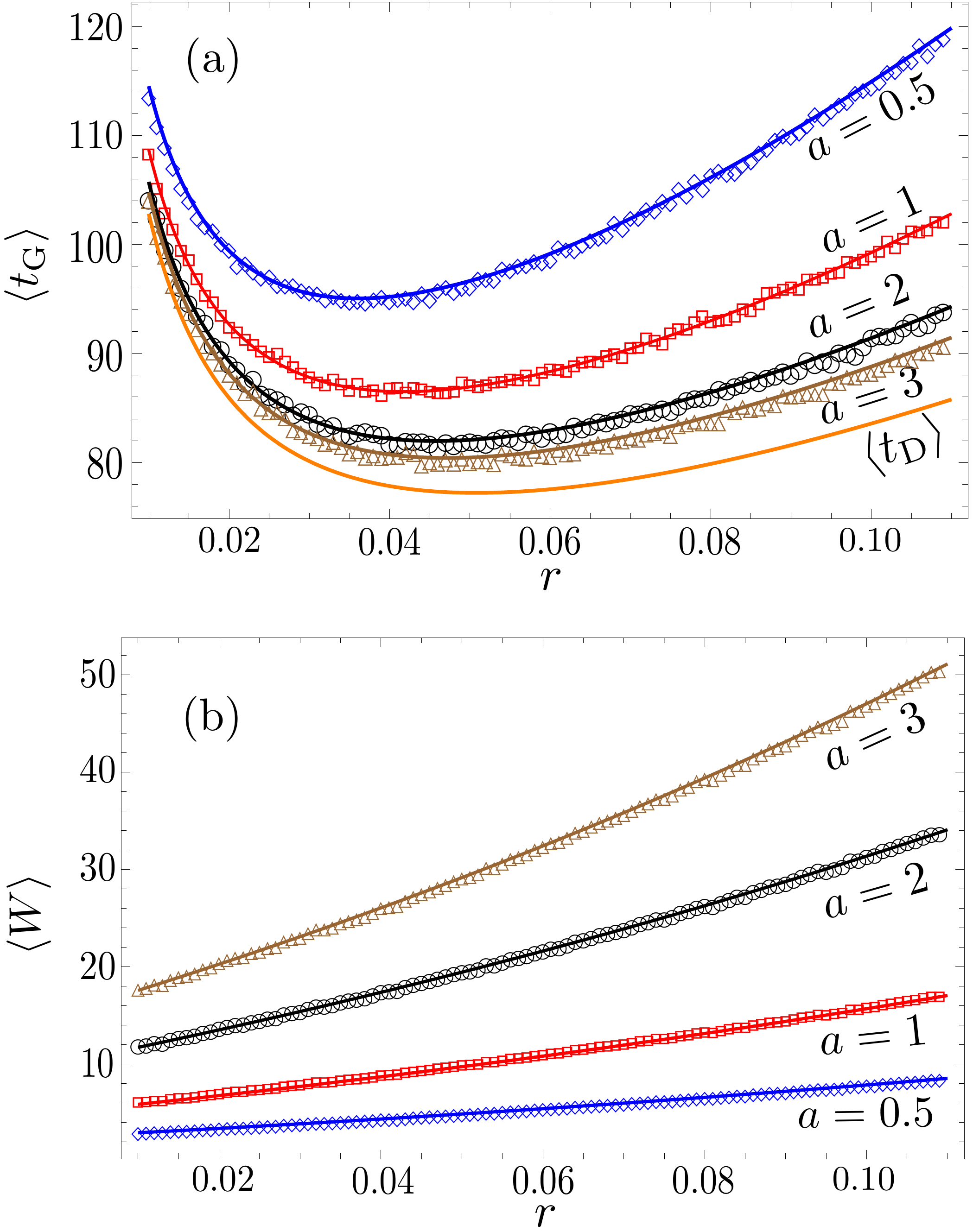}
	\caption{Analytic and numerical results for stochastic resetting with finite-time reset. Plots for (a) $\la t_{\rm G}\ra$ 
and (b) $\la W\ra$  as a function of $r$ for various values of $a$ with the parameters $x_0=x_{\rm R}=5$ and $D=D_{\rm R}=0.5$.  The solid curves of (a) and (b) represent the analytic results obtained from Eqs.~\eqref{fpt} and \eqref{work1}, respectively. All data points are obtained by averaging over $10^5$ trajectories.
	}
	\label{fig:FPT_W}
\end{figure}

Figure~\ref{fig:FPT_W}(a) displays the plot for $\langle t_{\rm G} \rangle$ versus $r$ for various values of $a$, which indicates that simulation results agree well with Eq.~\eqref{fpt}. As shown in the figure, $\langle t_{\rm G} \rangle$ is a non-monotonic function of $r$, thus, it is minimized at some optimal rate $r=r_{\rm G}^*$ which is the solution of $\partial_r \langle t_{\rm G} \rangle = 0$. As expected, $\langle t_{\rm G} \rangle$ approaches $\langle t_{\rm D} \rangle$ as $a$ increases. It has been demonstrated in Fig.~\ref{fig_supplementary} that  $r_{\rm G}^*$ and $\langle t_{\rm G}\rangle |_{r=r_{\rm G}^*}$ saturate the optimal rate $r_{\rm D}^*$ (solution of $\partial_r \langle t_{\rm D} \rangle = 0$) and the corresponding mean FPT $\langle t_{\rm D}\rangle |_{r=r_{\rm D}^*}$ for instantaneous resetting, respectively. Figure~\ref{fig:FPT_W}(b) displays the plot of $\langle W \rangle$ versus $r$. In contrast to $\langle t_{\rm G}\rangle $, $\langle W \rangle$ is a monotonically increasing function of both $r$ and $a$. Simulation data are in excellent agreement with Eq.~\eqref{work1}.

\begin{figure}
	\centering
	\includegraphics[width=0.45\textwidth]{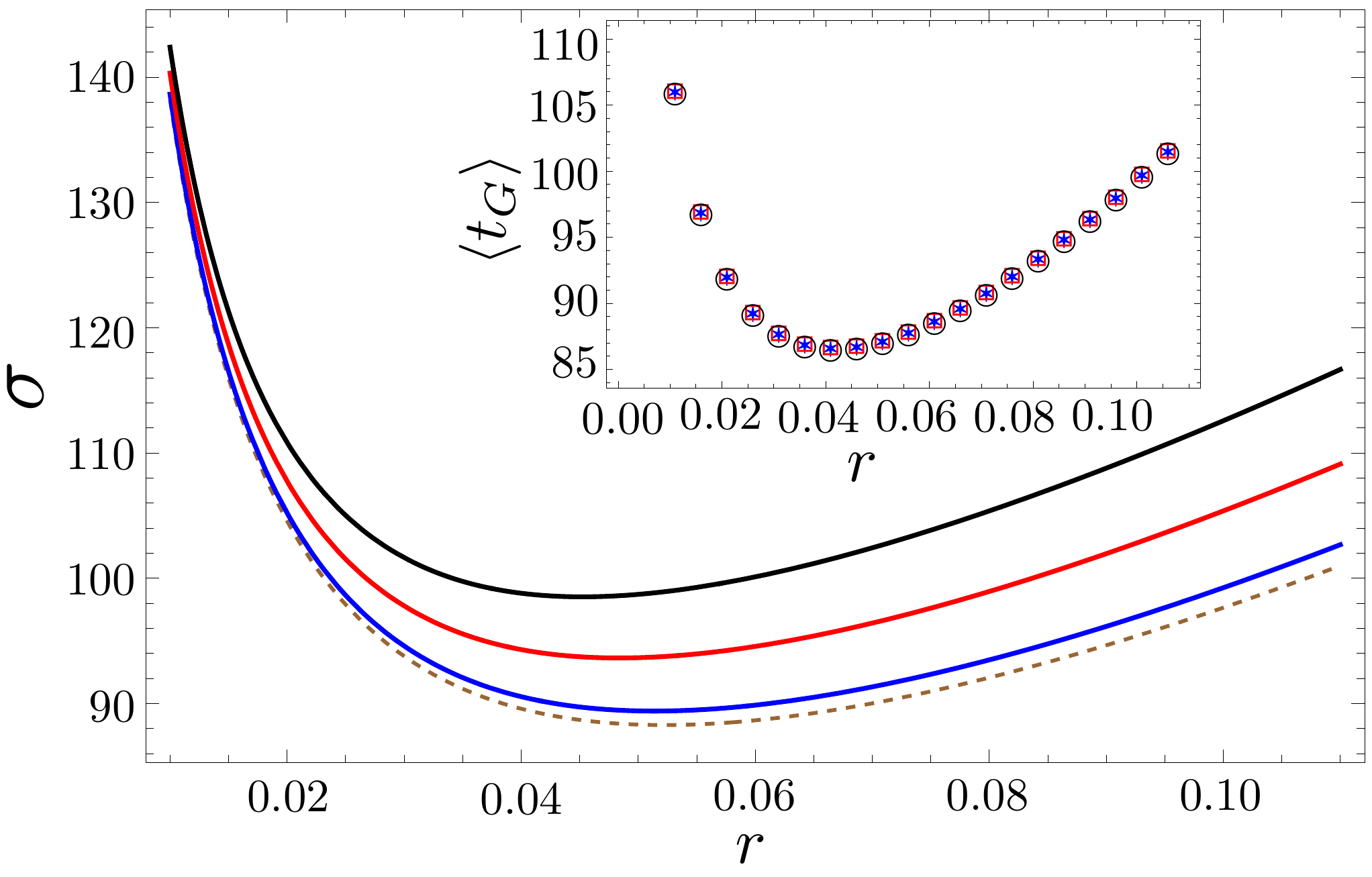}
	\caption{ Standard deviation of the global FPT as a function of the resetting rate $r$ for different values of $D_{\rm R}=0$ (Brown dashed), $D_{\rm R}=10$ (Blue), $50$ (Red), and $100$ (Black). (Inset) Mean FPT for different values of $D_{\rm R}$ which shows its invariance under $D_{\rm R}$ modulation unlike the standard deviation $\sigma$.  
	}
	\label{fig:std_dev_rel_fluc}
\end{figure}

We note that the mean FPT, Eq.~\eqref{fpt}, is independent of $D_{\rm R}$. In fact, Eq.~\eqref{fpt} is exactly the expression for the mean FPT as was obtained for the model with $D_{\rm R} = 0$ in~\cite{pal2020prr,bodrova2020pre1}. This is because the return time due to dragging with a constant velocity as was done in \cite{pal2020prr,bodrova2020pre1} is identical to the return time due to stochastic return (only in the mean level). However, the fluctuations in FPT and the role of higher order potentials should have different results compared to those with the deterministic reset dynamics. 
This point is emphasized in Fig.~\ref{fig:std_dev_rel_fluc}, which is the plot for the standard deviation of the global FPT $\sigma \equiv \sqrt{ \la t_{\rm G}^2\ra - \la t_{\rm G}\ra^2 }$ as a function of $r$ for various $D_{\rm R}$. The curves in the plot are evaluated from Eq.~\eqref{A_moments}. This plot clearly demonstrates that the fluctuation of the FPT depends on $D_{\rm R}$.

\section{Trade-off relation}   
\begin{figure}
	\centering
		\includegraphics[width=0.45\textwidth]{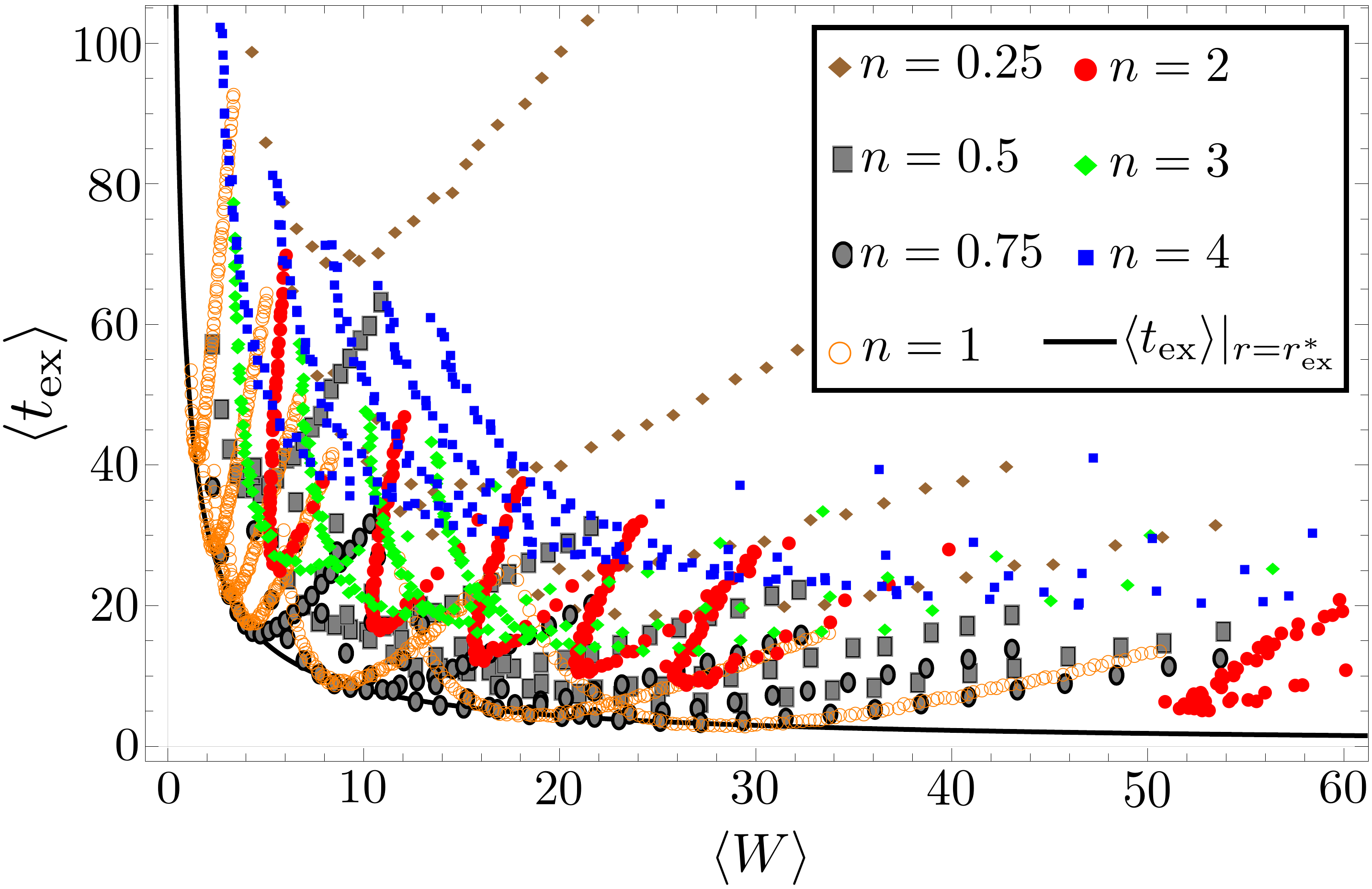}
		\caption{Trade-off relation between excess time and work, i.e., plot of $\la t_{\rm ex}\ra $ as a function of $\la W\ra$. Solid curve denotes the minimum excess time $\la t_{\rm ex}\ra |_{r=r_{\rm ex}^*}$ for a given work, derived for $n=1$. Data points are obtained from the simulation with various potential strength $a$, reset rate $r$, and potential exponent $n$. Each point is obtained by averaging over $10^5$ trajectories.    }
	\label{fig:tradeoff}
\end{figure}

It is evident that instantaneous resetting is not possible unless an infinite amount of work is provided. Thus, in order to address the physically meaningful question, ``what is the minimum FPT for a given cost?'',
we reformulate the mean FPT $\langle t_{\rm G}\rangle$ in terms of work $\langle W\rangle$ instead of the potential strength $a$, using Eqs.~\eqref{fpt} and \eqref{work1} as
\begin{align}
    \langle t_{\rm G}\rangle = \langle t_{\rm D} \rangle + \frac{1}{\alpha^2 \langle W \rangle } [2\sinh(\alpha x_{\rm R})- \alpha x_{\rm R}]^2.  \label{eq:tradeoff}
\end{align}
Equation~\eqref{eq:tradeoff} clearly shows the trade-off relation between mean FPT and average work: large work leads to small time, and vice versa. We introduce the notion of {\em excess} time $\la t_{\rm ex} \ra$ as 
the mean FPT in reference to the minimal mean FPT for instantaneous resetting, i.e., $\la t_{\rm ex} \ra \equiv 
\la t_{\rm G} \ra -\la t_{\rm D}\ra|_{r=r_{\rm D}^*}$.
By solving $\partial_r \la t_{\rm ex}\ra =0$ for fixed $\langle W\rangle$, we determine the resetting rate $r=r^*_{\rm ex}$ that minimizes $\langle t_{\rm ex}\ra$. 
The minimum excess time $\langle t_{\rm ex}\ra|_{r=r_{\rm ex}^*}$ is plotted against 
$\la W\ra$ in Fig.~\ref{fig:tradeoff}, along with simulation data for general (nonlinear) potentials of the form of Eq.~\eqref{eq:pot} with various values of $n$. It is noteworthy that, for any value of $n$, all data remain lower bounded by the minimum curve for $n=1$. This suggests that the minimum curve derived from the linear potential could serve as the lower bound of a universal time-cost trade-off relation for general finite-time stochastic resetting processes. Further study is necessary for elucidating the optimality of the bound.

Since our primary goal is to minimise mean FPT for given energy resources, it will be worthwhile to investigate other resetting strategy. In ~\cite{pal2016-sharp,pal2017}, it was shown that sharp resetting strategy - where the resetting is conducted stroboscopically i.e., after every fixed time interval $\tau_{\rm R}$ - can render the mean FPT globally optimized in the case of instantaneous resetting. It is thus natural to investigate the trade-off relation when sharp resetting is employed in the case of finite-time return. We have calculated the mean FPT and average work for a Brownian particle undergoing finite-time resetting process using sharp resetting protocol (details provided in Appendix D). In Fig.~\ref{fig:tradeoff1}, the FPT for sharp resetting protocol is plotted against average work done on the system for different $\tau_{\rm R}$. For comparison purpose, we also draw the optimal bound curve $\av{t_{\rm G}} |_{r=r_{\rm ex}^*}$ of the stochastic (Poissonian) resetting protocol, which is essentially     the same curve presented in Fig.~\ref{fig:tradeoff}. It is important to note that for some values of $\tau_{\rm R}$, the trade-off curve for sharp resetting protocol is well below the curve of $\av{t_{\rm G}} |_{r=r_{\rm ex}^*}$. This implies that for fixed energy resources, the mean FPT can be further lowered  using a sharp resetting protocol.
\begin{figure}
	\centering
		\includegraphics[width=0.5\textwidth]{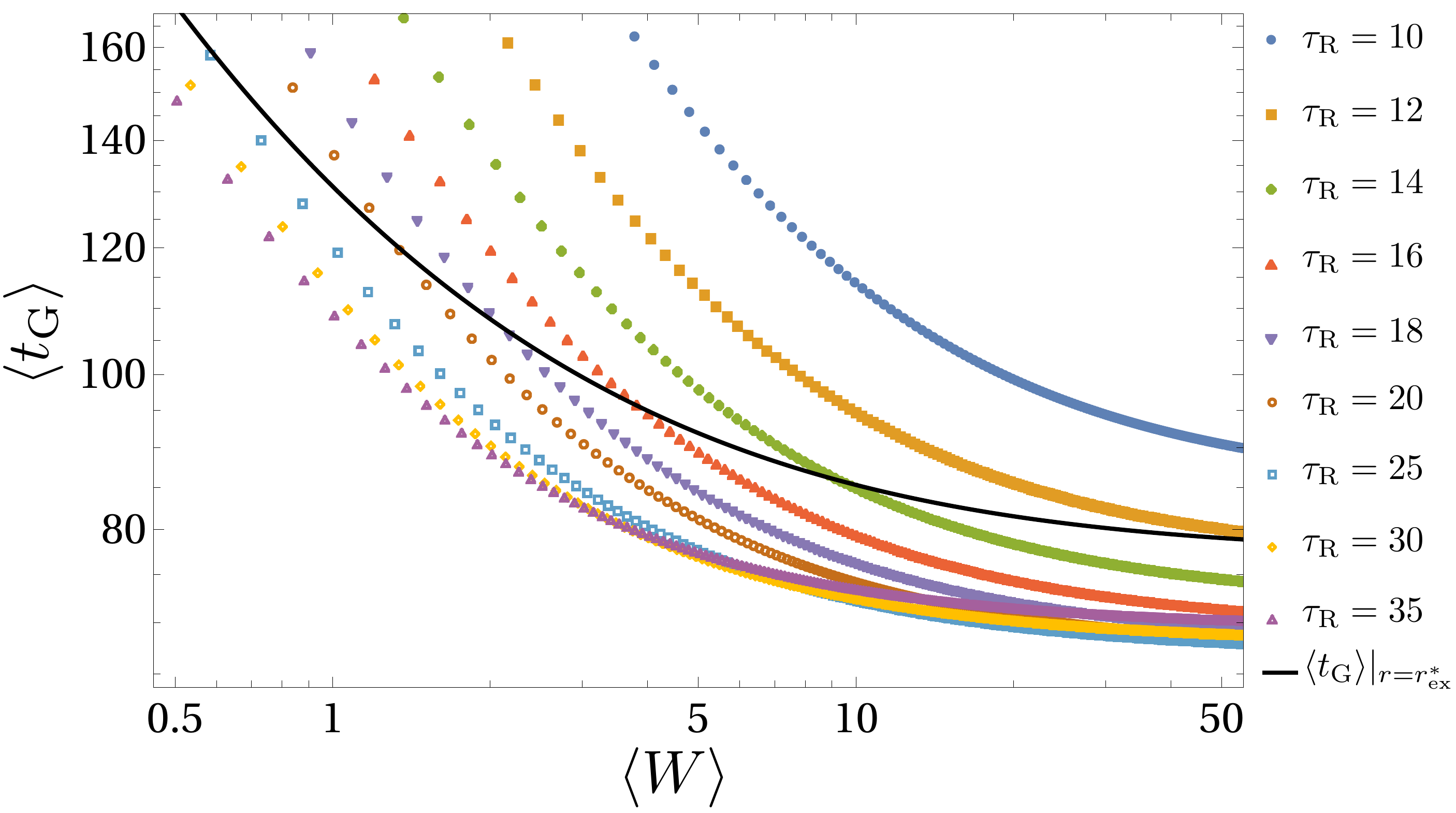}
 		\caption{Trade-off relation between mean FPT and average work. Solid curve denotes the minimum value of the global FPT $\av{t_{\rm G}} |_{r=r_{\rm ex}^*}$ for Poissonian (stochastic) resetting protocol. Data points are obtained by numerical calculation of mean FPT and $\av W$ for sharp resetting protocol for various $\tau_{\rm R}$. }
	\label{fig:tradeoff1}
\end{figure}

\section{Conclusion} 
In this study, we examined the thermodynamic cost and the first-passage time (FPT) of the stochastic resetting process, in which the reset is implemented using the trapping potential given by Eq.~\eqref{eq:pot}. We find a time-cost trade-off relation in stochastic resetting, where the minimum FPT can be decreased with increased work, and vice versa. 
Our result clearly demonstrates that while instantaneous resetting is always faster in target-searching, it requires an infinite cost, making it neither practical nor efficient from the viewpoint of energetics. 
The trade-off relation we found appears to be valid for a wide range of trapping potentials. Therefore, this trade-off relation could be used as a standard reference for investigating various processes accompanied with a finite-time stochastic resetting process, where the reset is not controllable but occurs at random time such as in biological systems. However, in the case where the reset is controllable such as in some artificial systems, the trade-off minima curve can be further lowered by using a different resetting strategy namely sharp resetting protocol.
Our results could lead to the construction of thermodynamically efficient searching strategies with finite energy resources, 
which could be especially useful in experimental studies of biophysical and single particle systems~\cite{BarZiv, Murugan,Bhaskaran, Hyeon, Chakrabarti,expt,expt-2} pertaining to finite-time stochastic resetting.

\section{Acknowledgement}
 We thank Changbong Hyun for many useful discussions and providing us the figures of proteins. The authors acknowledge Korea Institute for Advanced Study for providing computing resources [KIAS Center for Advanced Computation Linux Cluster System].  This research was supported by NRF Grant No.~2017R1D1A1B06035497 (H.P.), and individual KIAS Grants No.~PG064901 (J.S.L.), No.~PG085601 (P.S.P), and No.~QP013601 (H.P.) at Korea Institute for Advanced Study. AP gratefully acknowledges research support from the DST-SERB Start-up Research Grant Number SRG/2022/000080
and the DAE, Govt. of India. 

\begin{appendix}
\renewcommand{\theequation}{A\arabic{equation}}
\setcounter{equation}{0}
\section{Appendix A: Moment generating function of FPT in a single reset phase}
\label{appenA}
We consider a Brownian particle moving in an external potential $U(x)$ that is centered around the resetting position $x_{\rm R}$. The dynamics of the particle is described by the following Langevin equation
\begin{align}
\dot x=-\partial_xU(x)+\sqrt{2D_{\rm R}}\eta(t),
\end{align}
where $\eta(t)$ is a Gaussian white noise with zero mean and unit variance, and $D_{\rm R}$ is the diffusion constant for the reset phase. The particle starts at a position $x_1$ and we want to find the time when it reaches the position $x_{\rm R}$ for the first time. This is a typical reset phase scenario in a resetting dynamics with finite reset time.

The moment generating function of the FPT $t_{\rm R}$ to return to the resetting position can be written as
\begin{align}
Q_{\rm R}(p|x_1)=\int_0^{\infty} e^{-pt_{\rm R}}P_{\rm R}(t_{\rm R}|x_1)dt_{\rm R}=\la e^{-pt_{\rm R}}\ra,
\end{align}
where $P_{\rm R}(t_{\rm R}|x_1)$ is the probability density function of $t_{\rm R}$ given the particle starts the reset phase from the position $x_1$. Now we  divide FPT of reset process $t_{\rm R}$ into two parts: initial infinitesimal time $\Delta t$ and the remaining time $t_{\rm R} -\Delta t$. The position at time $t=\Delta t$ is given by $x_1'=x_1+[-\partial_xU(x)+ \sqrt{2D_{\rm R}}\eta(t)]\Delta t$. Therefore, the expression of $Q_{\rm R}(p|x_1)$ can be rewritten as 
\begin{widetext}
\begin{align}
Q_{\rm R}&(p|x_1)=\la e^{-p\Delta t}Q_{\rm R}(p|x'_1)\ra\nn\\
&\approx(1-p\Delta t) \la Q_{\rm R}(p|x_1-\partial_{x_1}U\Delta t+ \sqrt{2D_{\rm R}}\eta(0)\Delta t)\ra\nn\\
&\approx(1-p\Delta t) \bigg\la Q_{\rm R}(p|x_1)+\partial_{x_1}Q_{\rm R}(p|x_1)\{-\partial_{x_1}U+ \sqrt{2D_{\rm R}}\eta(0) \}\Delta t+\frac{1}{2}\partial^2_{x_1}Q_{\rm R}(p|x_1)\{-\partial_{x_1}U+ \sqrt{2D_{\rm R}}\eta(0)\}^2(\Delta t)^2 \bigg\ra\nn\\
&\approx Q_{\rm R}(p|x_1)-pQ_{\rm R}(p|x_1)\Delta t-(\partial_{x_1}U)\partial_{x_1}Q_{\rm R}(p|x_1)\Delta t+D_{\rm R}\partial^2_{x_1}Q_{\rm R}(p|x_1)\Delta t.\nn
\end{align}
\end{widetext}
Hence the backward differential equation of the moment generating function for the reset phase can be written as
\begin{align}
D_{\rm R}\partial^2_{x_1}Q_{\rm R}(p|x_1)-(\partial_{x_1}U)\partial_{x_1}Q_{\rm R}(p|x_1)-pQ_{\rm R}(p|x_1)=0.
\label{return_mgf1}
\end{align}
The boundary conditions for solving the above equation for $p>0$ are $Q_{\rm R}(p|x_1\rightarrow x_{\rm R})=1$ and $ Q_{\rm R}(p|x_1\rightarrow \pm \infty)=0$ with $ Q_{\rm R}(0|x_1)=1$. Since we are interested in the derivative of $Q_{\rm R}(p|x_1)$ with respect to $p$ at $p=0$, it is not necessary to consider the case for $p<0$.
We consider a linear trapping potential $U(x)=a|x_1-x_{\rm R}|$ $(a>0)$, yielding
\begin{align}
-\partial_{x_1}U&=-a, \hspace{1 cm} x_1\ge x_{\rm R}\nn\\
&=a. \hspace{1.3 cm} x_1< x_{\rm R}\nn
\end{align}
In the region $x_1\ge x_{\rm R}$, Eq.~\eqref{return_mgf1} is written as
\begin{align}
D_{\rm R}\partial^2_{x_1}Q_{\rm R}(p|x_1)-a\partial_{x_1}Q_{\rm R}(p|x_1)-pQ_{\rm R}(p|x_1)=0.
\end{align}
The solution of the above equation is given by
\begin{align}
Q_{\rm R}(p|x_1)=C_1e^{\lambda^+x_1}+C_2e^{\lambda^-x_1},
\end{align}
where $\lambda^{\pm}=\frac{a\pm\sqrt{a^2+4pD_{\rm R}}}{2D_{\rm R}}$. Using the boundary condition $Q_{\rm R}(p|x_1\rightarrow  \infty)=0\implies C_1=0$ and using $Q_{\rm R}(p|x_1\rightarrow x_{\rm R})=1 \implies C_2=e^{-\lambda^-x_{\rm R}}$. Hence, for $x_1\ge x_{\rm R}$,
\begin{align}
Q_{\rm R}(p|x_1)=\exp\bigg[-\frac{\sqrt{a^2+4pD_{\rm R}}-a}{2D_{\rm R}}(x_1-x_{\rm R})\bigg].
\label{return_mgf11}
\end{align}
In the region $x_1< x_{\rm R}$, Eq.~\eqref{return_mgf1} is written as
\begin{align}
D_{\rm R}\partial^2_{x_1}Q_{\rm R}(p|x_1)+a\partial_{x_1}Q_{\rm R}(p|x_1)-pQ_{\rm R}(p|x_1)=0. \label{eqA:reset_eq2}
\end{align}
The solution of the above equation~\eqref{eqA:reset_eq2} is given by
\begin{align}
Q_{\rm R}(p|x_1)=C_1e^{\lambda^+x_1}+C_2e^{\lambda^-x_1},
\end{align}
where $\lambda^{\pm}=\frac{-a\pm\sqrt{a^2+4pD_{\rm R}}}{2D_{\rm R}}$. Using the boundary condition $Q_{\rm R}(p|x_1\rightarrow  \infty)=0\implies C_2=0$ and using $Q_{\rm R}(p|x_1\rightarrow x_{\rm R})=1 \implies C_1=e^{-\lambda^+x_{\rm R}}$. Hence, for $x_1< x_{\rm R}$,
\begin{align}
Q_{\rm R}(p|x_1)=\exp\bigg[-\frac{\sqrt{a^2+4pD_{\rm R}}-a}{2D_{\rm R}}(x_{\rm R}-x_1)\bigg].
\label{return_mgf12}
\end{align}
Combining the two expressions in Eqs.~\eqref{return_mgf11} and \eqref{return_mgf12}, the moment generating function is
\begin{align}
Q_{\rm R}(p|x_1)=\exp\bigg[-\frac{\sqrt{a^2+4pD_{\rm R}}-a}{2D_{\rm R}}|x_1-x_{\rm R}|\bigg].
\label{return_mgf}
\end{align}

\renewcommand{\theequation}{B\arabic{equation}}
\renewcommand{\thefigure}{B\arabic{figure}}
\setcounter{equation}{0}
\setcounter{figure}{0}
\section{Appendix B: Calculation of the global FPT}

\begin{figure*}[!t]
	\centering
		\includegraphics[width=0.8\textwidth]{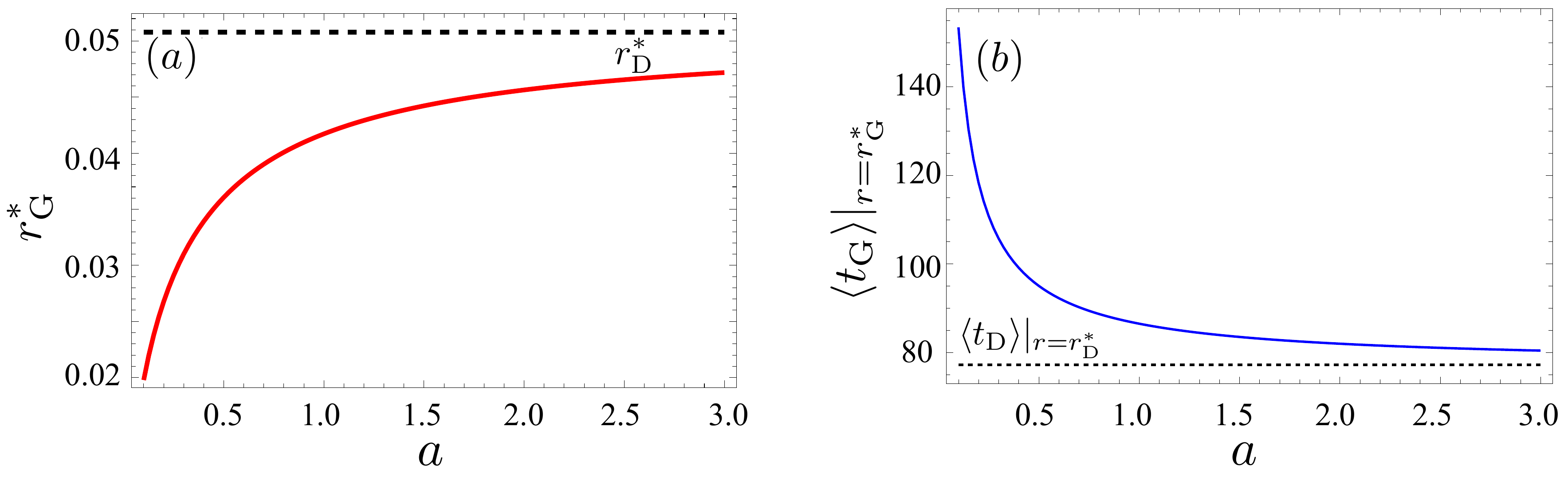}
		\caption{Analytical plots for linear trapping potential showing the convergence of the results of finite-time stochastic resetting to those of the instantaneous resetting process at large values of strength of the potential $a$.(a) Plot of optimal resetting rate $r_{\rm G}^*$ as a function of potential strength. (b) Plot of mean FPT at optimal resetting rate $r=r_{\rm G}^*$ as a function of potential strength. In both the plots the black dashed line represents the results of the instantaneous resetting.}
	\label{fig_supplementary}
\end{figure*}

The backward differential equation of $Q_{\rm A}(p|x_0)$ is given by
\begin{align}
&D\partial_{x_0}^2Q_{\rm A}(p|x_0)-(p+r) Q_{\rm A}(p|x_0)\nn\\
&\hspace{1.5 cm} +rQ_{\rm A}(p|x_{\rm R})Q_{\rm R}(p|x_0)=0 .
\label{bde_gt}
\end{align}
Using Eq.~\eqref{return_mgf}, the above equation can be rewritten  as 
\begin{align}
&D\partial_{x_0}^2Q_{\rm A}(p|x_0)-(p+r) Q_{\rm A}(p|x_0)\nn\\
&\hspace{1.5 cm}+r\exp[-\lambda(p)|x_0-x_{\rm R}|]Q_{\rm A}(p|x_{\rm R})=0,
\label{bde_gt1}
\end{align}
where $\lambda(p)=\frac{\sqrt{a^2+4pD_{\rm R}}-a}{2D_{\rm R}}$. The above equation~\eqref{bde_gt1} is solved in two regions namely, Region I: $x_0>x_{\rm R}$ and Region II: $x_0<x_{\rm R}$. In Region I, Eq.~\eqref{bde_gt1} can be written as
\begin{align}
&D\partial_{x_0}^2Q^I_{\rm A}(p|x_0)-(p+r) Q^I_{\rm A}(p|x_0)\nn\\
&\hspace{1.5 cm}+r\exp[-\lambda(p)(x_0-x_{\rm R})]Q_{\rm A}(p|x_{\rm R})=0. \label{eqA:B3}
\end{align}
The solution of Eq.~\eqref{eqA:B3} is 
\begin{align}
Q^I_{\rm A}(p|x_0)&=C_1e^{\mu(p)x_0}+C_2e^{-\mu(p)x_0}\nn\\
&\hspace{1.25 cm}-\nu(p)Q_{\rm A}(p|x_{\rm R})e^{-\lambda(p)(x_0-x_{\rm R})}, \label{eqA:B4}
\end{align}
where $\mu(p)=\sqrt{(p+r)/D}$ and $\nu(p)=r/[D\lambda^2(p)-(p+r)]$. In Region II, Eq.~\eqref{bde_gt1} can be written as
\begin{align}
&D\partial_{x_0}^2Q^{II}_{\rm A}(p|x_0)-(p+r) Q^{II}_{\rm A}(p|x_0)\nn\\
&\hspace{1.5 cm}+r\exp[-\lambda(p)(x_{\rm R}-x_0)]Q_{\rm A}(p|x_{\rm R})=0. \label{eqA:B5}
\end{align}
The solution of Eq.~\eqref{eqA:B5} is 
\begin{align}
Q^{II}_{\rm A}(p|x_0)&=C_3e^{\mu(p)x_0}+C_4e^{-\mu(p)x_0}\nn\\
&\hspace{1.25 cm}-\nu(p)Q_{\rm A}(p|x_{\rm R})e^{-\lambda(p)(x_{\rm R}-x_0)}. \label{eqA:B6}
\end{align}

The constants in the above expressions~\eqref{eqA:B4} and \eqref{eqA:B6} for $Q^I_{\rm A}(p|x_0)$ and $Q^{II}_{\rm A}(p|x_0)$ are determined by the following four boundary conditions: (i) $Q^I_{\rm A}(p|x_0\rightarrow\infty)=0$, 
(ii) $Q^{II}_{\rm A}(p|x_0\rightarrow 0)=1$, 
(iii) $Q^I_{\rm A}(p|x_0\rightarrow x_{\rm R}^+)=Q^{II}_{\rm A}(p|x_0\rightarrow x_{\rm R}^-)=Q_{\rm A}(p|x_{\rm R})$, and 
(iv) $\partial_{x_0}Q^I_{\rm A}(p|x_0)\bigg|_{x_0\rightarrow x_{\rm R}^+}=\partial_{x_0}Q^{II}_{\rm A}(p|x_0)\bigg|_{x_0\rightarrow x_{\rm R}^-}$. 

Boundary condition (i) suggests $C_1=0$. Using Boundary condition (ii),  we have 
\begin{align}
C_3+C_4-\nu(p)e^{-\lambda(p)x_{\rm R}}Q_{\rm A}(p|x_{\rm R})=1.
\label{const_eq1}
\end{align}
Using Boundary condition (iii), we get
\begin{widetext}
\begin{align}
C_2e^{-\mu(p)x_{\rm R}}-\nu(p)Q_{\rm A}(p|x_{\rm R})&=C_3e^{\mu(p)x_{\rm R}}+C_4e^{-\mu(p)x_{\rm R}}-\nu(p)Q_{\rm A}(p|x_{\rm R})\nn\\
\implies C_2&=C_3e^{2\mu(p)x_{\rm R}}+C_4.
\label{const_eq2}
\end{align}
Using Boundary condition (iv), we have
\begin{align}
-\mu(p)C_2e^{-\mu(p)x_{\rm R}}+\nu(p)\lambda(p)Q_{\rm A}(p|x_{\rm R})&=\mu(p)C_3e^{\mu(p)x_0}-\mu(p)C_4e^{-\mu(p)x_0}-\nu(p)\lambda(p)Q_{\rm A}(p|x_{\rm R})\nn\\
C_2e^{-\mu(p)x_{\rm R}}+C_3e^{\mu(p)x_0}-C_4e^{-\mu(p)x_0}&=\frac{2\nu(p)\lambda(p)}{\mu(p)}Q_{\rm A}(p|x_{\rm R})\nn\\
[C_3e^{2\mu(p)x_{\rm R}}+C_4]e^{-\mu(p)x_{\rm R}}+C_3e^{\mu(p)x_0}-C_4e^{-\mu(p)x_0}&=\frac{2\nu(p)\lambda(p)}{\mu(p)}Q_{\rm A}(p|x_{\rm R})\nn\\
2C_3e^{\mu(p)x_{\rm R}}&=\frac{2\nu(p)\lambda(p)}{\mu(p)}Q_{\rm A}(p|x_{\rm R})\nn\\
\implies  C_3&=\frac{\nu(p)\lambda(p)}{\mu(p)}Q_{\rm A}(p|x_{\rm R})e^{-\mu(p)x_{\rm R}}.
\label{const_eq3}
\end{align}
The expression of $C_4$ can be calculated by plugging Eq.~\eqref{const_eq3} into Eq.~\eqref{const_eq1}, 
\begin{align}
C_4=1+\nu(p)e^{-\lambda(p)x_{\rm R}}Q_{\rm A}(p|x_{\rm R})-\frac{\nu(p)\lambda(p)}{\mu(p)}Q_{\rm A}(p|x_{\rm R})e^{-\mu(p)x_{\rm R}}.
\label{const_eq4}
\end{align}
Using the expressions of $C_3$ and $C_4$ in Eq.~\eqref{const_eq2}, we have
\begin{align}
C_2&=\bigg[\frac{\nu(p)\lambda(p)}{\mu(p)}Q_{\rm A}(p|x_{\rm R})e^{-\mu(p)x_{\rm R}}\bigg]e^{2\mu(p)x_{\rm R}}+1+\nu(p)e^{-\lambda(p)x_{\rm R}}Q_{\rm A}(p|x_{\rm R})-\frac{\nu(p)\lambda(p)}{\mu(p)}Q_{\rm A}(p|x_{\rm R})e^{-\mu(p)x_{\rm R}}\nn\\
&=1+\nu(p)e^{-\lambda(p)x_{\rm R}}Q_{\rm A}(p|x_{\rm R})+\frac{\nu(p)\lambda(p)}{\mu(p)}Q_{\rm A}(p|x_{\rm R})\bigg[e^{\mu(p)x_{\rm R}}-e^{-\mu(p)x_{\rm R}}\bigg]\nn\\
&=1+\nu(p)e^{-\lambda(p)x_{\rm R}}Q_{\rm A}(p|x_{\rm R})+\frac{2\nu(p)\lambda(p)}{\mu(p)}Q_{\rm A}(p|x_{\rm R})\sinh[\mu(p)x_{\rm R}].
\end{align}
Hence the expression of the moment generating function in the region I is 
\begin{align}
Q^I_{\rm A}(p|x_0)&=e^{-\mu(p)x_0}+\nu(p)e^{-\lambda(p)x_{\rm R}}e^{-\mu(p)x_0}Q_{\rm A}(p|x_{\rm R})\nn\\
&\hspace{1 cm}+\frac{2\nu(p)\lambda(p)}{\mu(p)}Q_{\rm A}(p|x_{\rm R})\sinh[\mu(p)x_{\rm R}]e^{-\mu(p)x_0}-\nu(p)Q_{\rm A}(p|x_{\rm R})e^{-\lambda(p)(x_0-x_{\rm R})}. \label{eqA:B12}
\end{align}
The expression of $Q_{\rm A}(p|x_{\rm R})$ can be calculated by replacing $x_0$ with $x_{\rm R}$ in Eq.~\eqref{eqA:B12} as
\begin{align}
Q_{\rm A}(p|x_{\rm R})&=e^{-\mu(p)x_{\rm R}}+\bigg[\nu(p)e^{-[\lambda(p)+\mu(p)]x_{\rm R}}+\frac{2\nu(p)\lambda(p)}{\mu(p)}\sinh[\mu(p)x_{\rm R}]e^{-\mu(p)x_{\rm R}}-\nu(p)\bigg]Q_{\rm A}(p|x_{\rm R}),\nn\\
\implies Q_{\rm A}(p|x_{\rm R})&=\frac{e^{-\mu(p)x_{\rm R}}}{1+\nu(p)-\nu(p)e^{-[\lambda(p)+\mu(p)]x_{\rm R}}-\frac{2\nu(p)\lambda(p)}{\mu(p)}\sinh[\mu(p)x_{\rm R}]e^{-\mu(p)x_{\rm R}}}=\frac{e^{-\mu(p)x_{\rm R}}}{f_{\rm A}(p,x_{\rm R})}.
\end{align}
Considering the Brownian particle to be reset to its initial position i.e., $x_0=x_{\rm R}$, the FPT to reach the global target at the origin is 
\begin{align}
\la t_{\rm G}\ra&=-\partial_pQ_{\rm A}(p|x_{\rm R})\bigg|_{p\rightarrow 0},\nn\\
&=-\bigg[-x_{\rm R}\frac{e^{-\mu(p)x_{\rm R}}}{f_{\rm A}(p,x_{\rm R})}\partial_p\mu(p)-\frac{e^{-\mu(p)x_{\rm R}}}{f^2(p|x_{\rm R})}\partial_pf_{\rm A}(p,x_{\rm R})\bigg]\bigg|_{p\rightarrow 0},\nn\\
&=\bigg[x_{\rm R}Q_{\rm A}(p|x_{\rm R})\partial_p\mu(p)+e^{\mu(p)x_{\rm R}}Q^2_{\rm A}(p|x_{\rm R})\partial_pf_{\rm A}(p,x_{\rm R})\bigg]\bigg|_{p\rightarrow 0},\nn\\
&=x_{\rm R}Q_{\rm A}(0|x_{\rm R})\partial_p\mu(p)\bigg|_{p\rightarrow 0}+e^{\mu(0)x_{\rm R}}Q^2_{\rm A}(0|x_{\rm R})\partial_pf_{\rm A}(p,x_{\rm R})\bigg|_{p\rightarrow 0}.
\end{align}
In addition, we have 
\begin{align}
\mu(p)&=\sqrt{(p+r)/D}; \mu(0)=\sqrt{r/D}=\alpha;\partial_p\mu(p)=\frac{1}{\sqrt{D}}\frac{1}{2\sqrt{p+r}};\partial_p\mu(p)\bigg|_{p\rightarrow 0}=\frac{1}{2\sqrt{rD}}=\frac{1}{2D\alpha},\nn\\
\lambda(p)&=\frac{\sqrt{a^2+4pD_{\rm R}}-a}{2D_{\rm R}};\lambda(0)=0;\partial_p\lambda(p)=\frac{1}{\sqrt{a^2+4pD_{\rm R}}};\partial_p\lambda(p)\bigg|_{p\rightarrow 0}=\frac{1}{a},\nn\\
\nu(p)&=\frac{r}{D\lambda^2(p)-(p+r)};\nu(0)=\frac{r}{D\lambda^2(0)-r}=-1;\nn\\
\partial_p\nu(p)&=-\frac{r[2D\lambda(p)\partial_p\lambda(p)-1]}{[D\lambda^2(p)-(p+r)]^2};\nn\\
\partial_p\nu(p)\bigg|_{p\rightarrow 0}&=-\frac{r[2D\lambda(0)\partial_p\lambda(p)|_{p\rightarrow 0}-1]}{[D\lambda^2(0)-r]^2} = \frac{1}{r}.
\end{align}
The above expressions yield 
\begin{align}
\partial_pf_{\rm A}(p,x_{\rm R})\bigg|_{p\rightarrow 0}&=\partial_p\nu(p)\bigg|_{p\rightarrow 0}-\partial_p\nu(p)\bigg|_{p\rightarrow 0}e^{-[\lambda(0)+\mu(0)]x_{\rm R}}-\nu(0)x_{\rm R}e^{-[\lambda(0)+\mu(0)]x_{\rm R}}\bigg[\partial_p\lambda(p)|_{p\rightarrow 0}+\partial_p\mu(p)|_{p\rightarrow 0}\bigg]\nn\\
&\hspace{2 cm}-\frac{2\nu(0)}{\mu(0)}\sinh[\mu(0)x_{\rm R}]e^{-\mu(0)x_{\rm R}}\partial_p\lambda(p)\bigg|_{p\rightarrow 0},\nn\\
&=\frac{1}{r}-\frac{1}{r}e^{-\alpha x_{\rm R}}+(-1)x_{\rm R}e^{-\alpha x_{\rm R}}\bigg[\frac{1}{a}+\frac{1}{2\sqrt{rD}}\bigg]-\frac{2(-1)}{\alpha}\sinh[\alpha x_{\rm R}]e^{-\alpha x_{\rm R}}\frac{1}{a},\nn\\
&=\frac{1}{r}(1-e^{-\alpha x_{\rm R}})-x_{\rm R}e^{-\alpha x_{\rm R}}\bigg[\frac{1}{a}+\frac{1}{2D\alpha}\bigg]+\frac{2}{a\alpha}\sinh[\alpha x_{\rm R}]e^{-\alpha x_{\rm R}},\nn\\
Q_{\rm A}(0|x_{\rm R})&=\frac{e^{-\mu(0)x_{\rm R}}}{f(0,x_{\rm R})}\nn\\
&=\frac{e^{-\alpha x_{\rm R}}}{1+\nu(0)-\nu(0)e^{-[\lambda(0)+\mu(0)]x_{\rm R}}-\frac{2\nu(0)\lambda(0)}{\mu(0)}\sinh[\mu(0)x_{\rm R}]e^{-\mu(0)x_{\rm R}}}\nn\\
&=\frac{e^{-\alpha x_{\rm R}}}{1+(-1)-(-1)e^{-\alpha x_{\rm R}}}=1.
\end{align}
Therefore, the expression of the global mean FPT is
\begin{align}
\la t_{\rm G}\ra&=\frac{x_{\rm R}}{2D\alpha}+e^{\alpha x_{\rm R}}\bigg[\frac{1}{r}(1-e^{-\alpha x_{\rm R}})-x_{\rm R}e^{-\alpha x_{\rm R}}\bigg(\frac{1}{a}+\frac{1}{2D\alpha}\bigg)+\frac{2}{a\alpha}\sinh(\alpha x_{\rm R})e^{-\alpha x_{\rm R}}\bigg],\nn\\
&=\frac{x_{\rm R}}{2D\alpha}+\frac{1}{r}(e^{\alpha x_{\rm R}}-1)-\frac{x_{\rm R}}{a}-\frac{x_{\rm R}}{2D\alpha}+\frac{1}{a\alpha}\sinh(\alpha x_{\rm R}),\nn\\
&=\la t_{\rm D}\ra+\frac{1}{a\alpha}[2\sinh(\alpha x_{\rm R})-\alpha x_{\rm R}],
\end{align}
where $\la t_{\rm D}\ra=(e^{\alpha x_{\rm R}}-1)/r$ is the mean FPT to reach the global target at origin with instantaneous resetting.
\end{widetext}

\renewcommand{\theequation}{C\arabic{equation}}
\setcounter{equation}{0}
\section{Appendix C: Calculation of work}
The backward differential equation of $Q_{\rm C}(p|x_0)$ is given by
\begin{align}
&D\partial^2_{x_0}Q_{\rm C}(p|x_0)-rQ_{\rm C}(p|x_0)\nn\\
&\hspace{2.25 cm}+re^{-pw(x_0)}Q_{\rm C}(p|x_{\rm R})=0.
\label{eq_bf}
\end{align}
To calculate work during the whole process, we replace the weight function $w(x(t))$ with the trapping potential $U(x)$. Hence, for the linear potential, $w(x(t))=U(x)=a|x-x_{\rm R}|$ with $a>0$.  In this case Eq.~\eqref{eq_bf} becomes 
 \begin{align}
 &D\partial^2_{x_0}Q_{\rm C}(p|x_0)-rQ_{\rm C}(p|x_0)\nn\\
 &\hspace{2.25 cm}+re^{-pa|x_0-x_{\rm R}|}Q_{\rm C}(p|x_{\rm R})=0. 
 \label{work_eq1}
 \end{align}
 Equation~\eqref{work_eq1} is solved in two regions namely, Region I: $x_0>x_{\rm R}$ and Region II: $x_0<x_{\rm R}$. In Region I, Eq.~\eqref{work_eq1} can be written as
 \begin{align}
 D\partial^2_{x_0}Q^I_{\rm C}(p|x_0)-rQ^I_{\rm C}(p|x_0)+re^{-pa(x_0-x_{\rm R})}Q_{\rm C}(p|x_{\rm R})=0. \label{eqA:C3}
 \end{align}
 The solution of Eq.~\eqref{eqA:C3} is 
 \begin{align}
 Q^I_{\rm C}(p|x_0)&=C_1e^{\alpha x_0}+C_2e^{-\alpha x_0}\nn\\
 &\hspace{1.5 cm}-\gamma(p)e^{-pa(x_0-x_{\rm R})}Q_{\rm C}(p|x_{\rm R}),
 \end{align}
 where $\gamma(p)=r/(Dp^2a^2-2r)$. In the region II, Eq.~\eqref{work_eq1} can be rewritten as 
  \begin{align}
 D\partial^2_{x_0}Q^{II}_{\rm C}(p|x_0)-rQ^{II}_{\rm C}(p|x_0)+re^{-pa(x_{\rm R}-x_0)}Q_{\rm C}(p|x_{\rm R})=0. \label{eqA:C5}
 \end{align}
The solution of the above equation~\eqref{eqA:C5} is 
 \begin{align}
 Q^{II}_{\rm C}(p|x_0)&=C_3e^{\alpha x_0}+C_4e^{-\alpha x_0}\nn\\
 &\hspace{1.5 cm}-\gamma(p)e^{-pa(x_{\rm R}-x_0)}Q_{\rm C}(p|x_{\rm R}).
 \end{align}
Boundary conditions are: \\
(i) $Q^I_{\rm C}(p|x_0\rightarrow\infty)=0$, 
(ii) $Q^{II}_{\rm C}(p|x_0\rightarrow 0)=1$, 
(iii) $Q^I_{\rm C}(p|x_0\rightarrow x_{\rm R}^+)=Q^{II}_{\rm C}(p|x_0\rightarrow x_{\rm R}^-)=Q_{\rm C}(p|x_{\rm R})$, and \\
(iv) $\partial_{x_0}Q^I_{\rm C}(p|x_0)\bigg|_{x_0\rightarrow x_{\rm R}^+}=\partial_{x_0}Q^{II}_{\rm C}(p|x_0)\bigg|_{x_0\rightarrow x_{\rm R}^-}$. 

Boundary condition (i) suggests $C_1=0$. Using Boundary condition (ii), we have 
\begin{align}
C_3+C_4=1+\gamma(p)e^{-pax_{\rm R}}Q_{\rm C}(p|x_{\rm R}).
\end{align}
Using the boundary condition (iii), we get
\begin{widetext}
\begin{align}
C_2e^{-\alpha x_{\rm R}}-\gamma(p)Q_{\rm C}(p|x_{\rm R})&=C_3e^{\alpha x_{\rm R}}+C_4e^{-\alpha x_{\rm R}}-\gamma(p)Q_{\rm C}(p|x_{\rm R}),\nn\\
\implies C_2&=C_3e^{2\alpha x_{\rm R}}+C_4 .
\label{c2_eq}
\end{align}
Using the boundary condition (iv), we obtain
\begin{align}
-\alpha C_2e^{-\alpha x_{\rm R}}+pa\gamma(p)Q_{\rm C}(p|x_{\rm R})&=\alpha C_3e^{\alpha x_{\rm R}}-\alpha C_4e^{-\alpha x_{\rm R}}-pa\gamma(p)Q_{\rm C}(p|x_{\rm R}),\nn\\
\alpha \bigg[C_2e^{-\alpha x_{\rm R}}+C_3e^{\alpha x_{\rm R}}-C_4e^{-\alpha x_{\rm R}}\bigg]&=2pa\gamma(p)Q_{\rm C}(p|x_{\rm R}),\nn\\
C_3e^{\alpha x_{\rm R}}+C_4e^{-\alpha x_{\rm R}}+C_3e^{\alpha x_{\rm R}}-C_4e^{-\alpha x_{\rm R}}&=\frac{2pa}{\alpha }\gamma(p)Q_{\rm C}(p|x_{\rm R}),\nn\\
\implies C_3&=\frac{pa}{\alpha}\gamma(p)Q_{\rm C}(p|x_{\rm R})e^{-\alpha x_{\rm R}}. 
\end{align}
Using the expression in  Eq.~\eqref{c2_eq}, and substituting $x_0$ with $x_{\rm R}$ in the expression of $Q^I(p|x_0)$, one arrives at the following expression of $Q_{\rm C}(p|x_{\rm R})$:
\begin{align}
Q_{\rm C}(p|x_{\rm R})=\frac{e^{-\alpha x_{\rm R}}}{1+\gamma(p)-\gamma(p)e^{-(pa+\alpha)x_{\rm R}}-\frac{2pa}{\alpha}\gamma(p)\sinh(\alpha x_{\rm R})e^{-\alpha x_{\rm R}}}=\frac{e^{-\alpha x_{\rm R}}}{f_{\rm C}(p,x_{\rm R})}.
\end{align}
Considering the Brownian particle to be reset to its initial position, the average work done until the Brownian particle reaches the target for the first time is
\begin{align}
\la W\ra&=-\partial_pQ_{\rm C}(p|x_{\rm R})\bigg|_{p\rightarrow 0}=\frac{e^{-\alpha x_{\rm R}}}{f^2_{\rm C}(p,x_{\rm R})}\partial_pf_{\rm C}(p,x_{\rm R})\bigg|_{p\rightarrow 0},\nn\\
&=e^{\alpha x_{\rm R}}Q^2_{\rm C}(0|x_{\rm R})\partial_pf_{\rm C}(p,x_{\rm R})\bigg|_{p\rightarrow 0}.
\end{align}
Now, we have
\begin{align}
f_{\rm C}(p,x_{\rm R})&=1+\gamma(p)-\gamma(p)e^{-(pa+\alpha)x_{\rm R}}-\frac{2pa}{\alpha}\gamma(p)\sinh(\alpha x_{\rm R})e^{-\alpha x_{\rm R}},\nn\\
\partial_pf_{\rm C}(p,x_{\rm R})&=\partial_p\gamma(p)-\partial_p\gamma(p)e^{-(pa+\alpha)x_{\rm R}}+ax_{\rm R}\gamma(p)e^{-(pa+\alpha)x_{\rm R}}-\frac{2a}{\alpha}\gamma(p)\sinh(\alpha x_{\rm R})e^{-\alpha x_{\rm R}}\nn\\
&\hspace{3cm}-\frac{2pa}{\alpha}\partial_p\gamma(p)\sinh(\alpha x_{\rm R})e^{-\alpha x_{\rm R}},\nn\\
\partial_pf_{\rm C}(p,x_{\rm R})\bigg|_{p\rightarrow 0}&=\partial_p\gamma(p)\bigg|_{p\rightarrow 0}-\partial_p\gamma(p)\bigg|_{p\rightarrow 0}e^{-\alpha x_{\rm R}}+ax_{\rm R}\gamma(0)e^{-\alpha x_{\rm R}}-\frac{2a}{\alpha}\gamma(0)\sinh(\alpha x_{\rm R})e^{-\alpha x_{\rm R}},\nn\\
\gamma(p)&=\frac{r}{Da^2p^2-r}\implies \gamma(0)=-1, \partial_p\gamma(p)=-\frac{2Drpa^2}{(Da^2p^2-r)^2}\implies \partial_p\gamma(p)\bigg|_{p\rightarrow 0}=0,\nn\\
\therefore \partial_pf_{\rm C}(p,x_{\rm R})\bigg|_{p\rightarrow 0}&=-ax_{\rm R}e^{-\alpha x_{\rm R}}+\frac{2a}{\alpha}\sinh(\alpha x_{\rm R})e^{-\alpha x_{\rm R}}.
\end{align}
Hence, the final expression of the average work is
\begin{align}
\la W\ra&=e^{\alpha x_{\rm R}}Q^2(0|x_{\rm R})\bigg[-ax_{\rm R}e^{-\alpha x_{\rm R}}+\frac{2a}{\alpha}\sinh(\alpha x_{\rm R})e^{-\alpha x_{\rm R}}\bigg],\nn\\
&=\frac{a}{\alpha}[2\sinh(\alpha x_{\rm R})-\alpha x_{\rm R}].
\end{align}

\end{widetext}
\renewcommand{\theequation}{D\arabic{equation}}
\renewcommand{\thefigure}{D\arabic{figure}}
\setcounter{equation}{0}
\setcounter{figure}{0}
\section{Appendix D: Calculation of mean FPT and average work for sharp resetting protocol}
\begin{figure}
	\centering
	\includegraphics[width=0.45\textwidth]{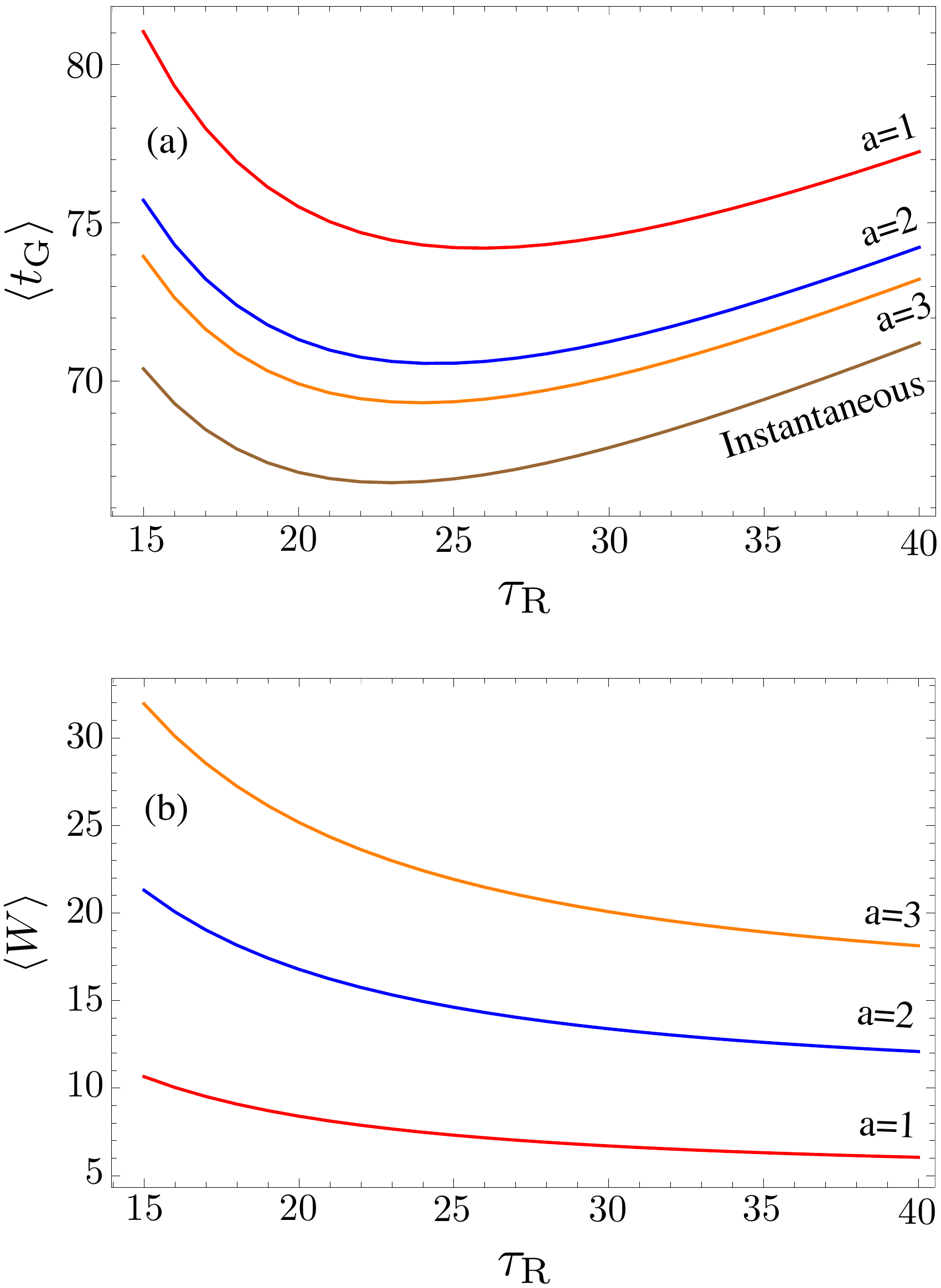}
	\caption{Numerical results for sharp resetting protocol. Plots for (a) $\la t_{\rm G}\ra$ 
and (b) $\la W\ra$  as a function of $\tau_{\rm R}$ for various values of $a$ with the parameters $x_0=x_{\rm R}=5$ and $D=D_{\rm R}=0.5$. }
	\label{fig:FPT_W_sharp}
\end{figure}

Consider a Brownian particle freely diffusing in one dimensional space. The particle can reach the target at a random time, say $T$, starting from an initial position $x_0$. However, resetting can occur before the particle finds the target resulting in resetting time $R<T$. In this case, the particle undergoes a return phase to the initial coordinate assisted by an external potential $U(x)$. Note that both  $T$ and $R$ can be sampled from arbitrary distribution. Let $x$ be the position of the particle when the reset phase starts and $\tau(x)$ be the mean time required to reach the resetting position $x_{\rm R}$ for the first time during the return phase. The mean global FPT to reach the target located at the origin then follows from \cite{pal2020prr}
\begin{align}
    \av{t_G}=\frac{\av{\text{min}(T,R)}}{Pr(T<R)}+\frac{\int_0^{\infty}dtf_R(t)\int dx \tau(x)G_0(x,t|x_0,0)}{Pr(T<R)},
    \label{eqd1}
\end{align}
where $f_R(t)$ is the probability density function of the resetting time. For example, in the case of stochastic resetting $f_R(t)=re^{-rt}$ and for sharp resetting $f_R(t)=\delta(t-\tau_R)$. The propagator $G_0(x,t|x_0,0)$ in Eq. (\ref{eqd1}) is the conditional probability density to find the particle at position $x$ at time $t$ given that it started at $x_0$ at time $t=0$, but in the presence of the target ~\cite{rednerbook}
\begin{align}
    G_0(x,t|x_0,0)=\frac{1}{\sqrt{4\pi Dt}}\bigg[e^{-\frac{(x-x_0)^2}{4Dt}}-e^{-\frac{(x+x_0)^2}{4Dt}}\bigg].
\end{align}
Finally, for a simple Brownian particle the mean reaching time $\tau(x)$ under the linear potential $U(x)=a|x-x_R|$ is given by
\begin{align}
    \tau(x)=-\partial_pQ_R(p|x)|_{p\rightarrow 0}=\frac{|x|}{a}.
\end{align}
Plugging all these expressions together into Eq.~\eqref{eqd1} does not yield a closed expression, hence we evaluate $\langle t_{\text{G}} \rangle$ numerically.  The result is plotted in Fig.~\ref{fig:FPT_W_sharp}(a).


Similar to the mean first passage time, one can construct a renewal equation for the work by noting that it depends only on the number of times the particle undergoes a reset phase. Following the method presented in Ref.~\cite{pal2020prr}, one can then write a renewal equation for the work as  
\begin{align}
    W&=0\hspace{2.5cm} \text{if}\hspace{.3cm} T<R\nn\\
    &=U(x)+W'\hspace{1cm}\text{if}\hspace{.3cm} R\le T, \label{eqA:D4}
\end{align}
where $W'$ is an independent and identically distributed copy of $W$ which again has the possibilities to accumulate zero or a finite quantity. The above expression~\eqref{eqA:D4} can be rewritten as 
\begin{align}
    W=I(R\le T)[U(x)+W'], \label{eqA:D5}
\end{align}
where $I(R\le T)$ is an indicator function which takes value 1 if $R\le T$ with probability $Pr(R\le T)$ and is zero otherwise. Taking expectations on the both sides of Eq.~\eqref{eqA:D5}, we have
\begin{align}
    \av W&=\av{I(R\le T)[U(x)+W']}\nn\\
    &=\av{I(R\le T)U(x)}+Pr(R\le T)\av{W'},
\end{align}
where in the last equality we have considered the fact that $W'$ is independent of $W$ and $\av{I(R\le T)}=Pr(R\le T)$. Finally, $\av W=\av{W'}$, since $W'$ is an independent and identically distributed copy of $W$.  Therefore a simple rearrangement leads to
\begin{align}
    \av W=\frac{\av{I(R\le T)U(x)}}{Pr(T<R)},
\end{align}
which can be computed as shown in Ref.~\cite{pal2020prr}. Following this, one finds
\begin{align}
    \av W=\frac{\int_0^{\infty}dtf_R(t)\int dx U(x)G_0(x,t|x_0,0)}{Pr(T<R)}.
    \label{appen_work_sharp}
\end{align}
For sharp resetting, we have calculated Eq.~\eqref{appen_work_sharp} numerically. The result is plotted in Fig.~\ref{fig:FPT_W_sharp}(b). Since $\tau_{\rm R}$ represents the time for diffusion phase, higher values of $\tau_{\rm R}$ implies lower number of reset phase and hence lower average work. We finally make a note that Eq.~\eqref{appen_work_sharp} is a very general expression that holds for arbitrary resetting time density, potential and underlying search process (and not limited to diffusion). \\   
\end{appendix}



\begin{thebibliography}{11}
\bibitem{mis} Despite the possibility of the particle touching the target in the reset phase, the reset dynamics continues prior to the reset point in our model. This likelihood is low, though, as the reset potential induces a directional motion of the particle to the reset point, away from the target. In the limit of a strong potential, this likelihood becomes negligible.
\end{thebibliography}
\end{document}